\documentclass[12pt,reqno]{amsart}

\usepackage[T1]{fontenc}
\usepackage[utf8]{inputenc}
\usepackage{lmodern}
\usepackage{microtype}
\usepackage{cite}
\usepackage{amsmath, amssymb, amsthm, amsfonts, mathrsfs}
\usepackage{mathtools}
\usepackage{bm}
\usepackage{bbm}
\usepackage{hyperref}
\usepackage{enumitem}
\usepackage{graphicx}
\usepackage{geometry}

\newcommand{\tz}{\tilde{\zeta}}

\makeatletter
\DeclareRobustCommand{\loplus}{\mathbin{\mathpalette\dog@lsemi{+}}}

\usepackage{float}
\usepackage{mathtools}
\newcommand{\defeq}{\vcentcolon=}

\newcommand{\dog@lsemi}[2]{\dog@semi{#1}{#2}{270,90}}
\newcommand{\dog@semi}[3]{%
  \begingroup
  \sbox\z@{$\m@th#1#2$}%
  \setlength{\unitlength}{\dimexpr\ht\z@+\dp\z@\relax}%
  \makebox[\wd\z@]{\raisebox{-\dp\z@}{%
    \begin{picture}(1,1)
    \linethickness{\variable@rule{#1}}
    \roundcap
    \put(0.5,0.5){\makebox(0,0){\raisebox{\dp\z@}{$\m@th#1#2$}}}
    \put(0.5,0.5){\arc[#3]{0.5}}
    \end{picture}%
  }}%
  \endgroup
}
\newcommand{\variable@rule}[1]{%
  \fontdimen8  
  \ifx#1\displaystyle\textfont3\else
    \ifx#1\textstyle\textfont3\else
      \ifx#1\scriptstyle\scriptfont3\else
        \scriptscriptfont3\relax
  \fi\fi\fi
}
\makeatother
\usepackage{tikz}
\usetikzlibrary{arrows.meta,positioning}

\usepackage{pict2e}
\usepackage{diagbox}

\newcommand{\beq}{\begin{eqnarray}}
\newcommand{\eeq}{\end{eqnarray}}
\newcommand{\beqn}{\begin{eqnarray}}
\newcommand{\eeqn}{\end{eqnarray}}

\newcommand{\spl}[1]{\mathrm{SL}(#1,\mathbb{R})}
\newcommand{\spla}[1]{\mathfrak{sl}\left(#1,\mathbb{R}\right)}

\usepackage{pict2e}

\newcommand{\updown}[2]{^{#1}_{\phantom{#1}#2}}

\newcommand{\chkM}{{\color{red} \,\checkmark\kern-5pt{}_{M}}}

\newcommand{\ee}{\end{equation}}
\newcommand{\bea}{\begin{eqnarray}}
\newcommand{\eea}{\end{eqnarray}}

\usepackage{BOONDOX-cal}
\usepackage{scalerel}
\usepackage{stackengine,wasysym}

\newcommand{\ad}[1]{\mathrm{ad}_{#1}}
\newcommand{\coad}[1]{\mathrm{ad}^*_{#1}}

\newcommand{\pairing}[2]{\langle #1\, , \, #2 \rangle}

\newenvironment{Align}{\begin{equation}
\begin{aligned}}
{\end{aligned}
\end{equation}}
\newenvironment{Align*}{\begin{equation*}
\begin{aligned}}
{\end{aligned}
\end{equation*}\par}
\usepackage{color}

\usepackage{layout}
\usepackage{hyperref}

\usepackage{cleveref}

\DeclareFontFamily{OT1}{rsfs}{}
\DeclareFontShape{OT1}{rsfs}{m}{n}{ <-7> rsfs5 <7-10> rsfs7 <10->rsfs10}{} 
\DeclareMathAlphabet{\mycal}{OT1}{rsfs}{m}{n}

\DeclareFontFamily{U}{MnSymbolC}{}
\DeclareSymbolFont{MnSyC}{U}{MnSymbolC}{m}{n}
\DeclareFontShape{U}{MnSymbolC}{m}{n}{
    <-6>  MnSymbolC5
   <6-7>  MnSymbolC6
   <7-8>  MnSymbolC7
   <8-9>  MnSymbolC8
   <9-10> MnSymbolC9
  <10-12> MnSymbolC10
  <12->   MnSymbolC12}{}
\DeclareMathSymbol{\intprod}{\mathbin}{MnSyC}{'270}
\usepackage{physics}
\usepackage[normalem]{ulem}
\hypersetup{
  colorlinks=true,
  linkcolor=blue!60!black,
  citecolor=blue!60!black,
  urlcolor=blue!60!black,
  pdftitle={Title},
  pdfauthor={First Author, Second Author}
}

\theoremstyle{plain}

\theoremstyle{definition}

\theoremstyle{remark}

\numberwithin{equation}{section}

\usepackage{cancel}

\title{Semi-Classical Limit of Quantum Gravity on Corners}

\author{Ludovic Varrin}
\address{National Centre for Nuclear Research, Pasteura 7, 02-093 Warsaw, Poland}
\email{ludovic.varrin@ncbj.gov.pl}

\date{\today}

\begin{document}
\begin{center}
\maketitle
\vspace{-1em}
\textsc{National Centre for Nuclear Research, Pasteura 7, 02-093 Warsaw, Poland}\\
\href{mailto:ludovic.varrin@ncbj.gov.pl}{ludovic.varrin@ncbj.gov.pl}
\end{center}
\begin{abstract}
We study quantum and classical systems associated with the quantum corner symmetry group
$\mathrm{QCS}=\widetilde{\mathrm{SL}}(2,\mathbb{R})\ltimes \mathrm{H}_3,$
which arises in the context of quantum gravity. We relate quantum observables---specified by representation-theoretic data---to their classical counterparts using generalized Perelomov coherent states and the framework of Berezin quantization. This procedure links abstract representation-theoretic input to geometric classical observables, such as area. We conclude by applying the formalism to static, spherically symmetric spacetimes admitting a horizon.

\end{abstract}

\vspace{.75cm}


% ---------------- Introduction ----------------
\section{Introduction}
The nature of quantum gravity has been an open problem in theoretical physics for the past century. One of the difficulties lies in the absence of a general quantization theorem and a reliance on quantization procedures that may or may not produce an adequate quantum version of the theory. The opposite direction, on the other hand, is much better understood. Given a quantum theory, it is possible to give a well defined notion of a classical limit---for example through the path-integral formulation of quantum mechanics \cite{Feynman:1942}. This is aligned with the understanding that the physical laws of the universe are quantum and that the associated classical theory emerges from an approximation of the more fundamental quantum version. The failure to successfully quantize gravity should therefore not be seen as an argument for the absence of such a theory but rather as a consequence of the particular description of the classical regime Einstein formulated in 1915 \cite{Einstein:1916vd}.\par
One might therefore wonder if anything can be said about the structure of quantum gravity without relying on the quantization of an underlying classical theory. One of the most promising way to address this problem lies in the use of \textit{symmetries}. Symmetries have been an essential tool in modern theoretical physics since Noether's theorems in the early 20th century \cite{Noether1918}\footnote{Coincidentally, Noether came up with the theorems while she was thinking about the new theory of general relativity.} which states that to any continuous symmetry of a system, there exists an associated conserved quantity. Latter, the symmetries of spacetime that drove Einstein to his special theory of relativity---the Poincaré symmetries--- were used by Wigner to classify all possible fundamental particles \cite{Wigner:1939cj}. His procedure gives the Hilbert space of particle physics without ever introducing a classical theory requiring some form of quantization.\par
So what are the symmetries of gravity? Any classical gravitational theory is based on diffeomorphism invariance---the invariance of the laws of physics under a change of coordinates on the spacetime manifold. However, this is nothing else than a redundancy in the description of the theory and, as such, their associated Noether charges vanish. The situation is different in the presence of boundaries. When spacetime boundaries are present---for example when thinking about a subregion of the spacetime--- some diffeomorphisms acquire a non-vanishing Noether charge on codimension-2 boundaries, corners \cite{Regge:1974zd,Wald:1999wa}. These charges further form an algebra of physical symmetries. In particular, it was shown in \cite{Ciambelli:2021vnn} that there exists a universal maximal algebra of physical symmetries for four-dimensional gravity: the universal corner symmetry (UCS) algebra
\begin{equation}
    \mathfrak{ucs} = \mathfrak{diff}(S) \loplus \qty(\mathfrak{gl}\qty(2,\mathbb{R})\loplus \mathbb{R}^2)^S,
\end{equation}
where $S$ denotes the corner and its presence in the exponents is to emphasize that there exists a copy of the algebra in the parenthesis at each point of the corner. The algebra is universal in the sense that it does not depend on the particular theory of gravity one might use. In particular, for corners situated at a finite distance in the bulk, the charges of Einstein- Hilbert theory realize the extended corner symmetry (ECS) algebra
\begin{equation}
    \mathfrak{ecs} = \mathfrak{diff}\qty(S) \loplus \qty(\spla{2} \loplus \mathbb{R}^2)^S.
\end{equation}
While these symmetry algebras emerge from classical consideration, one can study their representations with an agnostic view on their origin. This is at the heart of the corner proposal for quantum gravity, which states that a relevant set of quantum gravitational states should live in irreducible unitary representations of the corner symmetry group. This is the gravitational analogue of Wigner's classification of the Poincaré representations.\par
The program was initiated in \cite{Ciambelli:2024qgi} with the study of the two-dimensional version of the finite distance corner symmetry group
\begin{equation}
    \mathrm{ECS}_2 = \spl{2}\ltimes \mathbb{R}^2.
\end{equation}
The projective representations of the $\mathrm{ECS}_2$ are given by the standard representations of the maximally centrally extended version, the quantum corner symmetry (QCS) group
\begin{equation}
    \mathrm{QCS}= \widetilde{\spl{2}\ltimes} \mathrm{H}_3,
\end{equation}
where $H_3$ is the three dimensional Heisenberg group. These representations where given in \cite{Varrin:2024sxe,Neri:2025fsh} and will be described in the next section. The discrete series were used in \cite{Ciambelli:2025ztm,JKGLVtoappear} to describe local subsystems in two-dimensional quantum gravity---also seen as four-dimensional spherically symmetric quantum gravity--- and calculate their associated entanglement entropy. For some of the QCS coherent states that will be described later, the entanglement entropy of the subregion was shown to correspond to the area of the boundary of that subregion in the semi-classical limit, providing a symmetry based explanation for the emergence of the so-called area law.\par
This last point brings up an important question. Given that the Hilbert space was not obtained by the quantization of an underlying classical theory, how does one define a semi-classical limit for this formalism? The present work is dedicated to the mathematical formalism linking the representation theoretic objects of the quantum theory to the classical observables of a theory dynamically realizing the $\mathrm{ECS}_2$. The formalism consists in relating three spaces endowed with symplectic structures: the projective Hilbert space of the group representation, equipped with the Fubini–Study symplectic form; the coadjoint orbits of the group, equipped with the Kostant–Kirillov–Souriau symplectic form; and the classical phase space, equipped with the canonical symplectic form.
\par
The paper is organized as follows. In Section~\ref{sec:mathematicalbackground}, we present the necessary mathematical background, focusing on coadjoint orbits and their connection to the classical phase space. Section~\ref{sec:qcsdnyamicalsystem} is devoted to the application of this formalism to QCS dynamical systems. In the spirit of the introduction, we begin in Section~\ref{sec:quantumsystem} with the quantum system described by the representation theory of the QCS group, where coherent states provide a bridge to the coadjoint-orbit structure. Section~\ref{sec:classicalsystem} then introduces the classical QCS system and employs the moment-map construction to relate its dynamical variables to the quantum description through the same coadjoint orbits. Finally, in Section~\ref{sec:sssspacetimes}, we apply the procedure to the QCS dynamical system arising from static, spherically symmetric spacetimes and show that the area of the corner is directly related to the parameter labeling the representations. Concluding remarks are presented in Section~\ref{sec:conclusion}, and the proofs of key statements are relegated to the appendices.

% ---------------- Background ----------------
\section{Coadjoint Orbits and Moment Maps}\label{sec:mathematicalbackground}
We consider a Lie group $G$ with associated Lie algebra $\mathfrak{g}$, and a symplectic manifold $(\Gamma,\Omega_\Gamma)$---the phase space--- on which the group acts
\begin{equation}
    \triangleright_\Gamma : G\times \Gamma \longrightarrow\Gamma,\quad (g,\phi) \mapsto g\triangleright_\Gamma\phi.
\end{equation}
For each element of the Lie group $g\in G$ we define the map
\begin{equation}
    \Psi_g: G \longrightarrow \mathrm{Aut}\qty(G),
\end{equation}
which acts as the inner automorphism on group elements $h\in G$
\begin{equation}
    \Psi_g(h) = g h g^{-1}.
\end{equation}
The adjoint action is then defined as the differential of $\Psi_g$ at the identity element $e\in G$ 
\begin{equation}
    \mathrm{Ad}_g \defeq \qty(\dd \Psi_g)_e:\mathfrak{g}\longrightarrow \mathfrak{g}.
\end{equation}
The map
\begin{equation}
    \mathrm{Ad}: G\longrightarrow \mathrm{Aut}(\mathfrak{g}), \quad g\mapsto \mathrm{Ad}_g,
\end{equation}
is called the adjoint representation of the group $G$. Given a Lie algebra element $X\in\mathfrak{g}$, the derivative of the adjoint representation $\dd\qty(\mathrm{Ad})_e \defeq \mathrm{ad}$ produces the adjoint representation of the algebra
\begin{equation}
    \ad{X} : \mathfrak{g}\longrightarrow \mathfrak{g} \quad Y \longmapsto \ad{X}\qty(Y) \defeq \qty[X,Y] 
\end{equation}
The coadjoint action of the same algebra element on the coalgebra $\mathfrak{g}^*$ is defined by use of the pairing between vector and covectors
\begin{equation}
  \pairing{\coad{X}m}{Y} \defeq -\pairing{m}{\ad{X}Y} = \pairing{m}{\qty[Y,X]}.
\end{equation}
The coadjoint orbit through a point $m\in\mathfrak{g}^*$ is then defined as the submanifold
\begin{equation}
    \mathcal{O}_m = \qty{\mathrm{Ad}^*_g m \mid g\in G}.
\end{equation}
It will also be useful to define the infinitesimal version
\begin{equation}
\mathsf{O}_m = \qty{\coad{X}m \mid X\in \mathfrak{g}}.
\end{equation}
There are two key results concerning coadjoint orbits. The first being that each orbit is completely characterized by the stabilizer subalgebra
\begin{equation}
    \mathfrak{h}_m = \{\, X \in \mathfrak{g} \mid \coad{X}m = 0 \,\},
\end{equation}
as expressed by the isomorphisms
\begin{equation}\label{eq:coadjointisomorphism}
  \mathsf{O}_m \cong \mathfrak{g} / \mathfrak{h}_m, \quad \mathcal{O}_m \cong G/H_m,
\end{equation}
where $H_m \subset G$ is the stabilizer subgroup
\begin{equation}
H_m = \qty{h\in G\mid \mathrm{Ad}^*_h m = m} 
\end{equation}
whose Lie algebra is given by $\mathfrak{h}_m$.
The isomorphism \eqref{eq:coadjointisomorphism} follows directly from the bijectivity of the following orbit map 
\begin{equation}\label{eq:orbitmap}
    \tau: G/H_m \longrightarrow \mathcal{O}_m,\quad gH_m \mapsto \tau(gH_m) = \mathrm{Ad}^*_{g} m.
\end{equation}
The second key result is that any coadjoint orbit is naturally a symplectic manifold \cite{Souriau:1970,kostant1970,Kirillov2004}. This is seen as follows, for any element of the algebra $X$, define the fundamental vector field on the coadjoint orbits $X^\#$ by its action on a functional $f\in C^\infty(\mathcal{O}_m)$
\begin{equation}
    (X^\# f)(m) = \dv{t}\qty[\mathsf{f}\qty(\mathrm{Ad}^*_{\exp(t X)}m)]\eval_{t=0}.
\end{equation}
The symplectic structure is then given by the well-known Kostant--Kirillov--Souriau two-form, which is defined at the point $m\in\mathfrak{g}^*$ as
\begin{equation}
    \Omega^m_{\mathrm{KKS}}(X^\#,Y^\#) \defeq \pairing{m}{\qty[X,Y]}.
\end{equation}

The fundamental vector field is closely related to the linear coordinate functions on the coalgebra. These are defined as follows, given a basis of the algebra $X^a\in \mathfrak{g}$ satisfying
\begin{equation}
    \qty[X^a,X^b] = c^{ab}_c X^c,
\end{equation}
where $c^{ab}_c$ are the structure constants of the algebra, we define the coordinate function 
\begin{equation}\label{eq:coordinatefunctiondef}
    \chi^a: \mathfrak{g}^* \longrightarrow \mathbb{R},\quad \chi^a(m) = \pairing{m}{X^a}.
\end{equation}
It then follows that these are Hamiltonian functions for the fundamental vector field of the algebra basis
\begin{equation}\label{eq:coordinatehamiltonaian}
    \iota_{(X^a)^\#}\Omega_{KKS} = \dd \chi^a,
\end{equation}
where $\dd$ denotes the deRham differential on the coalgebra. The above equation also implies that the coordinate functions realize the algebra through the Poisson bracket induced by the KKS form
\begin{equation}
    \qty{\chi^a,\chi^b}_{\mathrm{KKS}}\defeq \iota_{(X^b)^\#}\iota_{(X^a)^\#}\Omega_{\mathrm{KKS}} = (X^b)^\#(\chi^a) = c^{ab}_c \chi^b.
\end{equation}
\par
On the field space side, we can also introduce a notion of fundamental vector field. We define the fundamental vector field associated with the algebra element $X\in\mathfrak{g}$ through its action on a function $\mathcal{f}\in C^{\infty}(\Gamma)$
\begin{equation}
    (\mathcal{X}\mathcal{f})(\phi) \defeq \dv{t}\qty[\mathcal{f}(\exp(t X)\triangleright_{\Gamma}\phi) ]\eval_{t=0}.
\end{equation}
The object relating the abstract coadjoint orbits to the phase space of the theory is the \textit{moment map} that we now describe. 
Let us now introduce a map
\begin{equation}\label{eq:momentmap}
    \mu: \Gamma \longrightarrow \mathfrak{g}^*,
\end{equation}
We call $\mu$ a moment map if its pairing with a vector is the Hamiltonian function for the fundamental vector field\footnote{Throughout this section, the differentials on both the coadjoint orbits and the phase space are denoted by $\dd$, and the corresponding contraction operators by $\iota$. The space on which these operators act will always be clear from the context.
}
\begin{equation}\label{eq:fundamentalhamiltonian}
    \iota_{\mathcal{X}} \Omega_\Gamma = \dd  \qty(\pairing{\mu}{X}) \defeq d\tilde{\mu}(X).
\end{equation}
where we defined the comoment map
\begin{equation}\label{eq:comomentmap}
    \tilde{\mu}: \mathfrak{g}\longrightarrow C^{\infty}\qty(\Gamma), \quad \tilde{\mu}\qty(X) \defeq \pairing{\mu}{X}.
\end{equation}
The comoment map further needs to satisfy the equivariance condition
\begin{equation}\label{eq:momentmapequivariance}
    \{\tilde{\mu}\qty(X),\tilde{\mu}\qty(Y)\}_\Gamma = \tilde{\mu}\qty([X,Y]).
\end{equation}
This map is what physicist usually call the Hamiltonian. Note that
\begin{equation}
    \tilde{\mu}(X^a) = \chi^a \circ \mu.
\end{equation}
One can therefore show (see Appendix \ref{ap:vectorfields}) that the push forward of the fundamental vector field on phase space is the fundamental vector field on the coadjoint orbits
\begin{equation}
    \mu_* \mathcal{X} = X^\#.
\end{equation}
\par

 The equivariance condition \eqref{eq:momentmapequivariance} may be obstructed by the presence of non-trivial cocycles in the algebra. This motivates the introduction of twisted coadjoint orbits and their assocaited twisted moment maps. The twisted coadjoint orbits are defined at the infinitesimal level as
\begin{equation}
 \mathsf{O}^c_m = \qty{\coad{X}m + c \sigma_X \mid X\in\mathfrak{g}}
\end{equation}
where $\sigma_X \in \mathfrak{g}^*$ is determined by the Lie algebra $2$-cocycle $\omega \in Z^2(\mathfrak{g},\mathbb{R})$, through
\begin{equation}
    \pairing{\sigma_X}{Y} = -\omega(X,Y), \qquad X,Y \in \mathfrak{g}.
\end{equation}
and where $c$ is the twisting parameter. In the following, we restrict our attention to the case where a single non-trivial cocycle exists, that is, when the second Lie algebra cohomology group
\begin{equation}
    H^2(\mathfrak{g},\mathbb{R}) \coloneqq Z^2(\mathfrak{g},\mathbb{R}) \big/ B^{2}\qty(\mathfrak{g},\mathbb{R}),
\end{equation}
with $B^{2}\qty(\mathfrak{g},\mathbb{R})$ denoting the space of 2-coboundaries—i.e., trivial cocycles—
is one-dimensional. The twisted coadjoint orbit through $m\in\mathfrak{g}^*$
is then isomorphic to the ordinary coadjoint orbits of the maximally centrally extended algebra at the point $(m,c)\in \tilde{\mathfrak{g}}^* = \mathfrak{g}^* \oplus \mathbb{R}$
\begin{equation}
    \mathcal{O}^c_m \cong \tilde{\mathcal{O}}_{(m,c)}.
\end{equation}

The twisting parameter is then nothing else than the value of the central element on the coadjoint orbits of $\tilde{\mathfrak{g}}$. Next, we define the twisted moment map
\begin{equation}
    \mu^c : \Gamma \longrightarrow \tilde{\mathfrak{g}}^*, \quad \mu^c(\phi) = \qty(\mu(\phi),c). 
\end{equation}

The twisted comoment map for a vector $(X,a)\in \tilde{\mathfrak{g}}$ is given by
\begin{equation}\label{eq:twistedcomap}
     \tilde{\mu}^c(X,a) \defeq \pairing{\mu^c}{(X,a)} = \mu\qty(X) + c a,
\end{equation}
and satisfies the twisted equivariance condition
\begin{equation}
    \qty{\tilde{\mu}^c\qty(X,a),\tilde{\mu}^c\qty(Y,b)}_\Gamma = \tilde{\mu}\qty([X,Y]) + c \,\omega(X,Y),
\end{equation}
which is the standard equivariance condition on the extended algebra $\tilde{\mathfrak{g}}$.
The correspondence between twisted coadjoint orbits of an algebra and the standard coadjoint orbits of its maximal central extension is the classical analogue of the relationship between projective representations of the algebra and ordinary representations of its maximal central extension in the quantum case.

%%%%%%%%%%%%%%%%%%%%%%%%%%%%%%%%%%%%%%%%%%%%%

% ---- Main Results ----
\section{QCS Dynamical System}\label{sec:qcsdnyamicalsystem}
QCS dynamical systems are theories that realize the QCS on their phase space. Quantum mechanically, this is expressed by the fact that the Hilbert space of the theory carries a projective unitary irreducible representation of the ECS$_2$. On the classical side, the Noether charge algebra realizes the ECS$_2$, and the twisted moment maps provide the corresponding realization of the QCS. The (twisted) coadjoint orbits introduced in the previous section serve as the bridge between the classical and quantum picture. We start with the quantum case by describing the representation theory of the QCS and connecting the structure with (twisted) coadjoint orbits with the use of generalized Perelomov coherent states \cite{APerelomov_1977}. Next, we treat the classical system and define the classical Casimir function on phase space. We then show that this function is independent of the choice of moment map. Finally, we show how this invariant Casimir can be used to associate classical data to the representation parameters.\par
In the following, we continue to use $\{X^a\}$ for a generic basis of the algebra $\mathfrak{qcs}$, while also employing the explicit basis $\{J^a{}_b,\, P^a,\, Z\mid a,b=0,1\}$, where the generators $J^a{}_b$ span the subalgebra $\mathfrak{sl}(2,\mathbb{R})$ and $\{P^a,\, Z\}$ span the Heisenberg algebra
\begin{equation}\label{eq:qcsalgebra}
    [P^a, P^b] = \epsilon^{ab} Z, 
    \qquad [P^c, J^a{}_b] = \delta^c_b P^a - \tfrac{1}{2}\,\delta^a_b P^c, 
    \qquad [J^a{}_b, J^c{}_d] = \delta^c_b J^a{}_d - \delta^a_d J^c{}_b.
\end{equation}
The element $Z$ is central and commutes with all other generators.
\subsection{Quantum system}\label{sec:quantumsystem}
The representations of the QCS are given by tensor-product representations of $\spl{2}$ and the Weil–Schrödinger representations~\cite{Varrin:2024sxe,Neri:2025fsh}
\begin{equation}
    \pi_{(\kappa,c)} = \pi^{\spl{2}}_{\kappa} \otimes \pi^{\mathrm{W.S.}}_{c}.
\end{equation}
In particular, the discrete series of QCS can be realized on the Hilbert space
\begin{equation}
    \mathcal{H}^{(\kappa,c)}_{\mathrm{QCS}}
    = \bigl\{\,\ket{(\kappa,c);n,k} = \ket{\kappa;n}\otimes \ket{c;k}\,\bigr\},
\end{equation}
where $\{\ket{\kappa;n}\,:\, n\in\mathbb{N}\}$ denotes the lowest- (highest-) weight representation of the positive (negative) discrete series of $\widetilde{\mathrm{SL}}(2,\mathbb{R})$~\cite{Kitaev:2017hnr,Sun:2021thf}, and $\{\ket{c;k},\, k \in \mathbb{Z}^+\}$ ($\{\ket{c;k},\, k \in \mathbb{Z}^-\}$) denotes the lowest (highest) weight Weil–Schrödinger representation of the Schrödinger group with positive (negative) central element $c$~\cite{howe_1973,Berndt1998ElementsOT}. 
The Heisenberg operators act by raising or lowering the $k$ index, while the $\spl{2}$ ladder generators raise and lower the $n$ index by $\pm1$ and the $k$ index by $\pm2$.
The central element acts trivially on any state
\begin{equation}
    \dd \pi_{(\kappa,c)}(Z)\ket{(\kappa,c);n,k} = c\,\ket{(\kappa,c);n,k},
\end{equation}
where we denote the induced algebra representation by $\dd \pi$. There exists another Casimir operatorin the universal enveloping algebra $U[\mathfrak{qcs}]$\footnote{To give a precise meaning of the element $Z^{-1}$ one can look at the quotient
\begin{equation}
    U[\mathfrak{qcs}\times \mathbb{R}]/I_Z,
\end{equation}
where the additional factor of $\mathbb{R}$ is generated by $Z^{-1}$, $U[\mathfrak{g}]$ denotes the universal enveloping algebra and $I_Z$ is the ideal generated by $Z^{-1}Z-1$.
}
\begin{equation}\label{eq:casimirqcs}
    \mathcal{C}_{\mathrm{QCS}} =  \kappa\updown{a}{b}\,\updown{c}{d}   J^b_a J^d_c + \frac{1}{2Z} \epsilon_{ac} P^c J^a_b P^b,
\end{equation}
with
\begin{equation}
    \kappa\updown{0}{0}\, \updown{0}{0} = 1,\quad \kappa\updown{0}{1}\,\updown{1}{0}= \frac14, \quad \kappa\updown{1}{0}\,\updown{0}{1} = \frac34.
\end{equation}
It acts on irreducible representation as
\begin{equation}
    \dd \pi_{(\kappa,c)}(\mathcal{C}_{\mathrm{QCS}})\ket{(\kappa,c);n,k}
    = \qty(\kappa(\kappa - 1) + \tfrac{3}{16})\ket{(\kappa,c);n,k}.
\end{equation}
For more details on these representations, see~\cite{Berndt1998ElementsOT,Varrin:2024sxe,Neri:2025fsh}.
\par
The generalized Perelomov coherent states for the QCS discrete series were first introduced in \cite{Ciambelli:2025ztm}. They are constructed on the reference state $\ket{(\kappa,c);\Omega} = \ket{(\kappa,c);n=0,k=0}$ by first identifying the stabilizer subgroup
\begin{equation}
    H = \qty{g\in QCS \mid \pi_{(s,c)}(g)\ket{(\kappa,c);\Omega}= e^{i\theta_g}\ket{(\kappa,c);\Omega}},
\end{equation}
for some phase $\theta_g$. From this point onward we will always work in a fixed representation $\pi_{(s,c)}$ and drop the representation label for clarity. They will be reintroduced when necessary. The isotropy subgroup of the QCS for the vacuum state is given by \cite{Ciambelli:2025ztm}
\begin{equation}
    H = U(1) \times Z(\mathrm{H}_3).
\end{equation}
where $Z(\mathrm{H}_3) \cong \mathbb{R}$ is the center of the Heisenberg group. Consider the principle $H$ bundle 
\begin{equation}
    G \xlongrightarrow{\pi_G} G/H,\quad \pi_G(g) = \qty[g].
\end{equation}
Given a section $\sigma: G/H \longrightarrow G$---a choice of representative---, we define the displacement operator
\begin{equation}\label{eq:displacementoperator}
    U_\sigma([g]) = \pi(\sigma(\qty[g])).
\end{equation}
The generalized Perelomov coherent states are then defined as
\begin{equation}
    \ket{\sigma(g)} = U_\sigma(\qty[g]) \ket{\Omega}.
\end{equation}
A different choice of section only changes the coherent state by a phase factor.
Finally, we note that the coset is isomorphic to the coadjoint orbits
\begin{equation}
    QCS/H \cong \spl{2}/U(1) \times \mathrm{H}_3/Z(\mathrm{H}_3) = \mathbb{D} \times \mathbb{R}^2.
\end{equation}
We can therefore parametrize elements of the coset using coordinates on the coadjoint orbits $\zeta\in \mathbb{D}, \alpha \in \mathbb{C}$. A convenient choice of representative is then given by
\begin{equation}\label{eq:perelomovsection}
    g_{\zeta,\alpha} = \exp(\frac{1}{\sqrt{c}}\qty(\alpha P^1 - \bar{\alpha}P^0))\exp(c_\zeta J_+ -\bar{c_\zeta}J_-),
\end{equation}
where for $\zeta= r e^{i\theta}$, we defined $c_\zeta = \tanh^{-1}(r) e^{i\theta}$ and
\begin{Align}
    J_+ &= -\frac12(J^1_0 + J^0_1) + iJ^0_0,\\
    J_+ &= -\frac12(J^1_0 + J^0_1) - iJ^0_0.
\end{Align}
The associated coherent states are then given by
\begin{equation}\label{eq:coherentstatedef}
    \ket{\zeta,\alpha} = \pi(g_{\zeta,\alpha})\ket{\Omega}. 
\end{equation}
Thus, the QCS coherent states are a tensor product of Perelomov $\spl{2}$ coherent states \cite{APerelomov_1977} and squeezed Glauber coherent states \cite{Glauber1963a,Glauber1963b,Stoler1970,Yuen1976} with squeezing parameter $c_\zeta$
\begin{equation}
    \ket{\zeta,\alpha} = \ket{\zeta} \otimes \ket{\alpha_{c_\zeta}}.
\end{equation}
These quantum states, parametrized by coordinates on the orbit, are the key to relating the representation-theoretic structure to classical observables. To this end, we define the \textit{Berezin symbol} as the expectation value of algebra operators in coherent states.
\begin{equation}\label{eq:berezinsymbols}
    l_{(\zeta,\alpha)} : \mathfrak{qcs} \longrightarrow \mathbb{R}, 
    \qquad 
    l_{(\zeta,\alpha)}\qty(X) \defeq
    \mel{\zeta,\alpha}{\dd\pi(X)}{\zeta,\alpha}.
\end{equation}
Given an element of the algebra $X\in \mathfrak{qcs}$, we also define its associated \textit{Berezin function} as
\begin{equation}
    l^X : \tilde{\mathcal{O}} \longrightarrow \mathbb{R}, \quad l^X(\zeta,\alpha) \defeq l_{(\zeta,\alpha)}\qty(X).
\end{equation}
Since the coadjoint orbits---together with moment maps---form the phase space of the classical system, the Berezin functions can naturally be interpreted as classical observables. 
The associated Berezin Poisson bracket is then defined by
\begin{equation}\label{eq:berezinbracket}
    \{ l^X, l^Y \}_B
    \coloneqq 
    \frac{1}{i\hbar}
    l^{\qty[X,Y]}.
\end{equation}
Denote a basis of the $\mathfrak{qcs}$ by $X^a, a=1,...,6$ and the associated structure constants by $c^{ab}_{c}$. The algebra representation obey
\begin{equation}
    \qty[\dd \pi(X^a),\dd \pi\qty(X^b)] = i\hbar \,c^{ab}_c\dd \pi\qty(X^c).
\end{equation}
In this basis, the Berezin bracket then reads
\begin{equation}
    \qty{l^{X^a},l^{X^b}}_B = c^{ab}_c l^{X^c},
\end{equation}
which reproduces the Lie--Poisson structure on the coadjoint orbit. The Berezin functions equipped with the bracket \eqref{eq:berezinbracket}, are reminiscent of the coordinate functions \eqref{eq:coordinatefunctiondef} equipped with the KKS bracket. This is not a coincidence as in the case where the integral orbits are Kähler, one can show (see Appendix \ref{ap:berezinequalcoordinate}) that the Berezin functions are hamiltonian functions of the fundamental vector fields with respect to the KKS form
\begin{equation}\label{eq:berezinishamiltonian}
   \iota_{X^{\#}}\Omega_{\mathrm{KKS}} = \dd l^{X}.
\end{equation}
Note that this directly implies
\begin{equation}
    \qty{l^X,l^Y}_{B} = \qty{l^X,l^Y}_{\mathrm{KKS}}.
\end{equation}
Equation~\eqref{eq:berezinishamiltonian} implies that the coordinate functions and the Berezin functions differ, at most, by a constant. Since we have shown above that all choices of moment map are equivalent, we may set this constant to zero without loss of generality.
This brings us to the main result of the paper: the classical observables of a QCS system are obtained as the coherent-state expectation values of the quantized operators associated with the algebra generators. 
More precisely, the (twisted) comoment maps can be expressed as the composition of the Berezin function with the moment map
\begin{equation}
    \tilde{\mu}^c(X) = l^X \circ \mu^c.
\end{equation}
where, in the above expression, $\mu^c$ is the particular choice of twisted moment map that lands on the orbit $\tilde{\mathcal{O}}^{(s,c)}$. This establishes the correspondence between the classical observables, expressed as moment maps, and the quantum operators of the $\mathrm{QCS}$ algebra.\par
While this correspondence is exact for linear coordinate functions, the situation is more subtle for higher-order operators. This point is particularly important, since quadratic operators encode quantum fluctuations
\begin{equation}
    \expval{\qty(\Delta X)^2}_{\zeta,\alpha} \defeq \expval{\dd \pi\qty(X)^2}_{\zeta,\alpha} - \expval{\dd\pi\qty(X)}^2_{\zeta,\alpha}.
\end{equation}
 We start by extending the definition of the Berezin symbol to quadratic polynomials in the enveloping algebra $U[\mathfrak{qcs}]$
\begin{equation}
    l_{(\zeta,\alpha)}(XY) = \mel{\zeta,\alpha}{\dd \pi(X)\dd \pi(Y)}{\zeta,\alpha},
\end{equation}
The associated Berezin function then admits the asymptotic expansion \cite{Berezin1975,FABerezin_1975,RAWNSLEY199045,Bordemann:1993zv,Karabegov1996,Schlichenmaier:2000lmh,Charles2003} 
\begin{equation}\label{eq:Berezinexpansion}
  l^{XY} = l^X l^Y + \hbar \mathcal{h}^{\bar{a}b}\partial_{\bar{a}} l^X \partial_b l^Y + \mathcal{O}\qty(\hbar^2),
\end{equation}
where $\mathcal{h}^{\bar{a}b}$ are the components of the inverse Kähler metric. The full asymptotic expansion yields the Berezin–Toeplitz star product, which furnishes a deformation quantization of the underlying Kähler manifold.\par
In order to better understand the Kähler structure on the orbits, we make the following change of coordinates
\begin{equation}\label{eq:changeofcoordinates}
    \tilde{\chi}^a_b = \chi^a_b - \frac{1}{2 c} \epsilon_{bc}\chi^c \chi^a.
\end{equation}
In these coordinates, the special linear and Heisenberg sectors decouple
\begin{equation}
    \qty{\tilde{\chi}^a_b,\tilde{\chi}^c_d} = \delta^c_b \tilde{\chi}^a_d - \delta^a_d \tilde{\chi}^c_b, \quad \qty{\tilde{\chi}^a_b,\chi^c} = 0,
\end{equation}
and the KKS form reduces to the direct sum of the respective $\spl{2}$ and Heisenberg KKS forms~\cite{Neri:2025fsh}. In the associated holomorphic coordinates\footnote{The isomorphism between the upper sheet hyperboloid in $\mathfrak{sl}\qty(2,\mathbb{R})^*$ and the unit disk is given by
\begin{equation}
    (\tilde{\chi}^0_1,\tilde{\chi}^1_0,\tilde{\chi}^0_0) \longmapsto \kappa\qty(\frac{\qty(1-\tilde{\zeta})\qty(1-\bar{\tilde{\zeta}})}{1-\tilde{\zeta}\bar{\tz}}, \frac{(1+\tz)(1+\bar{\tz})}{ \tz \bar{\tz}-1},\frac{i(\tz-\bar{\tz})}{1-\tz \bar{\tz}}),
\end{equation}
and the isomorphism between the orbits of the Heisenberg group with the complex plane is given by
\begin{equation}
    \qty(\chi^1,\chi^2)\longmapsto \frac{\sqrt{c}}{2}\bigl(\alpha + \bar{\alpha},i(\bar{\alpha}-\alpha)\bigr).
\end{equation}
}, the QCS Kähler form can thus be written
\begin{equation}
    \Omega_{\mathrm{QCS}} = -2i\kappa \frac{\dd \tilde{\zeta} \wedge \dd \bar{\tilde{\zeta}}}{\qty(1 - \tilde{\zeta}\bar{\tilde{\zeta}})^2} + \frac{i}{2}\dd \alpha \wedge \dd \bar{\alpha}.
\end{equation}
In other words, the Kähler metric for the QCS is given by
\begin{equation}\label{eq:kahlermetric}
    \mathcal{h} = -2 \kappa \frac{ \dd \tilde{\zeta} \dd \bar{\tilde{\zeta}}}{\qty(1 - \tilde{\zeta}\bar{\tilde{\zeta}})^2} + \frac12\dd \alpha \dd \bar{\alpha}. 
\end{equation}
Finally we rewrite the above in the standard form by collecting the Kähler indices into a vector $k^a$ with $k^0 = \tilde{\zeta}$ and $k^1 = \alpha$
\begin{equation}
    \mathcal{h} = \mathcal{h}_{a\bar{b}} \dd k^a \dd \bar{k}^b.
\end{equation}
The inverse Kähler metric coefficients entering the expansion \eqref{eq:Berezinexpansion} are then defined by the usual property
\begin{equation}
    \mathcal{h}^{\bar{c}a}\mathcal{h}_{b\bar{c}}= \delta^a_b, \quad \mathcal{h}_{a\bar{c}}\mathcal{h}^{\bar{c}b} = \delta^b_a. 
\end{equation}
The first order correction in \eqref{eq:Berezinexpansion} can be decomposed into its symmetric and antisymmetric parts
\begin{equation}
    \hbar \mathcal{h}^{\bar{a}b}\partial_{\bar a}l^X \partial_b l^Y = C_1^{\mathrm{S}}\qty(X,Y) + C_1^{\mathrm{AS}}\qty(X,Y).
\end{equation}
 It is easy to see that the antisymmetric part of the first order corrections is the KKS bracket 
 \begin{equation}
     C^{\mathrm{AS}}_1\qty(X,Y) = \frac{\hbar}{2i}\qty{l^X,l^Y}_{\mathrm{KKS}}.
 \end{equation}
The symmetric part
 \begin{equation}
     C^{\mathrm{S}}_1\qty(X,Y) = \frac12\mathcal{h}^{\bar{a}b}\qty(\partial_{\bar{a}}l^X \partial_b l^Y + \partial_{\bar{a}}l^Y\partial_b l^X),
 \end{equation}
controls the quantum fluctuation in the classical limit
\begin{equation}
    \expval{\qty(\Delta X)^2}_{\zeta,\alpha} \defeq \expval{\dd \pi\qty(X)^2}_{\zeta,\alpha} - \expval{\dd\pi\qty(X)}^2_{\zeta,\alpha} = C^{\mathrm{S}}_1\qty(X,X) + \mathcal{O}\qty(\hbar^2).
\end{equation}
The above formula therefore represents the leading-order contribution, in the semi-classical limit, of the quantum fluctuations in terms of classical data. \par
This concludes our analysis of the quantum QCS system. The representation-theoretic data has been successfully linked with the coadjoint orbit structure. As explained in Section~\ref{sec:mathematicalbackground}, these are related to the classical system through the moment map. This connection will be the subject of the next section.
\subsection{Classical system and classical limit}\label{sec:classicalsystem}
Given a classical field theory, the covariant phase space formalism associates a symplectic structure on the field space of solutions of the equation of motion \cite{Crnkovic:1986ex,Crnkovic:1987tz,Lee:1990nz}. This allows one to associate Noether charges to symplectomorphism of the theory through Hamilton's equation \eqref{eq:fundamentalhamiltonian}\footnote{In principle, Hamilton's equation might be non integrable. There exists however several ways to remedy the issue \cite{Wald:1999wa,Barnich:2007bf,Donnelly:2016auv,Ciambelli:2021nmv,Klinger:2023qna}}. These are the comoment maps discussed earlier. As mentioned, there might exists obstruction to the equivariance condition. A classical QCS system is therefore a classical theory whose twisted comoment maps satisfy the QCS equivariance condition. As mentioned in the introduction, this is the case of JT gravity and spherically symmetric four-dimensional Einstein-Hilbert theory for finite distance corners.
\par
The Casimir function on the coalgebra $\mathfrak{qcs}^*$ is defined by
\begin{equation}\label{eq:casimirfunctionQCS}
    c_{QCS} =  \kappa\updown{a}{b}\,\updown{c}{d}  \chi^b_a \chi^d_c + \frac{1}{2 \chi_z} \epsilon_{ac}\chi^c \chi^a_b \chi^b.
\end{equation}
The fact that this function is constant on each individual orbit follows directly from its invariance under the coadjoint action. The second term in \eqref{eq:casimirqcs} is the Casimir function of the $\mathfrak{ecs}$ \cite{Ciambelli:2022cfr}. The pointwise product of functions being Abelian futher allows to write the Casimir function as
\begin{equation}\label{eq:casimirfunction}
    c_{\mathrm{QCS}} =   \, c_{\spl{2}} + \frac{1}{\chi_z}c_{ECS},
\end{equation}
where we defined the Casimir functions
\begin{align}
    c_{\spl{2}} &= \frac12 \chi^a_b \chi^b_a,\\
    c_{ECS} &=  \frac12 \epsilon_{ac}\chi^c \chi^a_b \chi^b.
\end{align}
\par
Let us now consider a field theory with phase space $\Gamma$ and symplectic form $\Omega_\Gamma$. The system is a QCS dynamical system if there exists (twisted) comoment maps
\begin{Align}\label{eq:momentmaplambda}
    \tilde{\mu}(J^a_b) \defeq \mu^a_b,\\
    \tilde{\mu}(P^a)  \defeq \mu^a\\
    \tilde{\mu}(Z)\defeq \mu^z,
\end{Align}
whose field space bracket defined by $\Omega_\Gamma$ reproduces the $\mathfrak{qcs}$ algebra \eqref{eq:qcsalgebra}.
We then define the classical Casimir as the function on phase space given by
\begin{Align}\label{eq:classicalcasimir}
    C_{\mathrm{QCS}}^{\mathrm{cl}} &\defeq  \kappa\updown{a}{b}\,\updown{c}{d} \,\mu^b_a \mu^d_c + \frac{1}{2\mu^z} \epsilon_{ac}\mu^c \mu^a_b \mu^b\\
    &=  C^{\mathrm{cl}}_{\spl{2}} + \frac{1}{c}C^\mathrm{cl}_{\mathrm{ECS}},
\end{Align}
Note that the classical Casimir and the Casimir functions are related by
\begin{equation}
  C^{\mathrm{cl}}_{\mathrm{QCS}} =  \mu^{c*}(c_{QCS}),\quad C^{\mathrm{cl}}_{\spl{2}} =  \mu^{c*}(c_{\spl{2}}),\quad C^{\mathrm{cl}}_{\mathrm{ECS}} =  \mu^{c*}(c_{ECS}).
\end{equation}
\par
We now address the following question: if one selects a different set of moment maps \eqref{eq:momentmaplambda}, how does the classical Casimir function transform?  
Since any admissible choice must satisfy the equivariance condition \eqref{eq:momentmapequivariance}, the allowed transformations of the moment maps are characterized by a matrix $A \in \mathrm{GL}\qty(2,\mathbb{R})$ and take the form
\begin{align}
    \mu'^a_b &= A^a_c \mu^c_d \qty(A^{-1})^d_b,\\
    \mu'^a &= A^a_b \mu^b,\\
    \mu'^z &= \det(A)\,\mu^z.
\end{align}
It can then be shown (see Appendix~\ref{ap:casimir}) that the classical Casimir function is invariant under such transformation. While this results follows from Schur's lemma in the semi-simple case, no complete results are known in the general case. This result provides a precise classical counterpart to the quantum Casimir operator. 
\par
While the definition \eqref{eq:casimirfunctionQCS} gives a well defined notion of the classical Casimir, the presence of the coordinate function associated with the central element in the denominator makes it inconvenient to discuss the semiclassical limit where we take the twisting parameter to zero. A more convenient choice is
\begin{equation}\label{eq:modifiedcasimir}
    c'_{\mathrm{QCS}} = \chi_z\, c_{\mathrm{QCS}}. 
\end{equation}
Denoting a point in the dual algebra by its component in terms of the dual basis of \eqref{eq:qcsalgebra} $\tilde{p} = (j^a_b,p^a,c)\in \mathfrak{qcs}^*$\footnote{To be precise: 
\begin{equation}
    \chi^a_b(p) \defeq j^a_b, \quad \chi^a(p) \defeq p^a, \quad \chi_z(p) \defeq c.    
\end{equation}}, the value of this alternative Casimir function on the orbit of that point is given by
\begin{equation}
    c'_{\mathrm{QCS}}\eval_{\tilde{\mathcal{O}}_{\tilde{p}}} = c \, \kappa\updown{a}{b}\, \updown{c}{d}j_a^b j_c^d + \frac12 \epsilon_{ac}p^c j_b^a p^b.
\end{equation}
In the semiclassical limit $c\mapsto 0$, the above function therefore reduces to the value of the ECS Casimir function the ECS coadjoint orbit of the point $p = (j^a_b,p^a)\in \mathfrak{ecs}^*$. When using this Casimir, one should keep in mind that it is defined only up to an overall multiplicative constant: different choices of moment map lead to different values of this prefactor.\par
We can now make precise the correspondence between quantum and classical data for QCS dynamical systems. On the classical side, the moment map~\eqref{eq:momentmap} assigns to each field configuration in the classical phase space a point in $\mathfrak{ecs}^*$. On the quantum side, evaluating Berezin symbols \eqref{eq:berezinsymbols} yields a comoment map for the algebra $\mathfrak{qcs}$, and thus determines a point on the corresponding QCS coadjoint orbit by interpreting the Berezin functions as coordinate functions on the orbit. In the untwisting limit $c\to 0$, this construction reduces to a point on an (untwisted) ECS coadjoint orbit, which can then be identified with the point selected by the classical moment map.\par

In the next section, we will carry out this procedure to relate abstract representation-theoretic data to geometric objects in a classical spacetime, focusing on the case of static, spherically symmetric spacetimes.
\section{Spherically Symmetric Static Spacetimes}\label{sec:sssspacetimes}
In this section, we give a explicit example of the correspondence between classical and quantum data for a specific QCS system; spherically symmetric static (SSS) spacetimes. In the context of the corner proposal for gravity \cite{Ciambelli:2022vot,Ciambelli:2023bmn}, the classical phase space $\Gamma$ consists of the gravitational field and its derivatives evaluated at the corner under scrutiny. A moment map $\mu: \Gamma \rightarrow \mathfrak{g}^*$ then associates to a specific field configuration, i.e. a specific geometry, a point on the dual algebra of the symmetry group. Field variations then moves this point along a coadjoint orbit through the equivariance condition
\begin{equation}\label{eq:equivariancecondition}
    \mu \circ \delta_X  = \mathrm{ad}^*_X \circ \mu.
\end{equation}
If $\mathfrak{g} = \mathfrak{ecs}$, the twisted moment maps realize the $\mathfrak{qcs}$ algebra and $\Gamma$ provides a QCS dynamical system. In that case, the Noether charge associated with a diffeomorphism $\xi \in \mathrm{T}M$ can be written
\begin{equation}
    H_\xi = \qty(\xi^\mu{}_{,\nu}N^{\nu}_\mu + t^{\mu}\xi_\mu)\eval_{S},
\end{equation}
where $N^\mu_\nu, t^\mu \in C^\infty(\Gamma)$, $N^\mu_\nu$ is traceless and where the corner is situated at the origin of the coordinate system $x^\mu(S) = 0$. The functional $N^a_b, t^a$ are thus generated by the first order expansion of diffeomorphisms around the corner \cite{Ciambelli:2021vnn}. Within the extended phase space formalism \cite{Ciambelli:2021nmv}, these vector fields are Hamiltonian vector fields and the Poisson bracket on $\Gamma$ realizes the $\mathfrak{ecs}$ algebra
\begin{equation}
    \qty{N^\mu_\nu,N^\rho_\sigma}_\Gamma = \delta^\rho_\nu N^\mu_\sigma - \delta^\mu_\sigma N^\rho_\nu, \quad \qty{t^\rho,N^\mu_\sigma}_\Gamma = \delta^\rho_\nu t^\mu - \frac12 \delta^\mu_\nu t^\rho, \quad \qty{t^\mu,t^\nu}_\Gamma = 0.
\end{equation}
\par
Let us now discuss the explicit example of SSS spacetimes.
For simplicity, we work in dimensionless units and reinstate them at the end of the calculation. We start by introducing the Hermitian basis of the QCS $\qty{D,H,K,X,P,Z}$, with commutation relations
\begin{Align}
    \qty[D,H] &= i H, \quad \qty[D,K] = - i K, \quad \qty[K,H] = 2 i D\\
    \qty[D,X] &= -\frac{i}{2}X ,\quad \qty[D,P] = \frac{i}{2} P, \quad \qty[H,X] = -i P,\\
    \qty[K,P] &= i X, \quad \qty[X,P] = i Z,\quad \qty[K,X] = \qty[H,P] = 0.
\end{Align}
We give the Berezin symbols for a general coherent states $\zeta = r e^{i\theta}\in \mathbb{D} , \alpha \in \mathbb{C}$\footnote{We are working in the lower hyperboloïd $l_{(\zeta,\alpha)}(h)<0$ and $l_{(\zeta,\alpha)(k)}< 0$. This corresponds to the negative discrete series of the special linear group $D^-_{\kappa}$.}
\begin{align}
    l_{(\zeta,\alpha)}(D) &= -\frac{r \sin\theta}{1-r^2}\qty(2 \kappa + \frac12) + \frac12 \Im(\alpha) \Re(\alpha),\label{eq:lD}\\
    l_{(\zeta,\alpha)}(H) &= -\frac{(\kappa-\frac14) \Delta}{(1-r^2)} + \Im(\alpha)^2,\label{eq:lH}\\
    l_{(\zeta,\alpha)}(K) &= -\frac{(\kappa -\frac14)\tilde{\Delta}}{1-r^2}+\frac{\Re(\alpha)^2}{4},\label{eq:lK}\\
    l_{(\zeta,\alpha)}(X) &= \frac{\sqrt{c}\Re(\alpha)}{\sqrt{2}},\label{eq:lX}\\
    l_{\zeta,\alpha}(P) &= \sqrt{2 c}\Im(\alpha),\label{eq:lP}
\end{align}
where we defined $\Delta = 1 -2 r \cos\theta + r^2$ and $\tilde{\Delta} = 1 + 2 r \cos\theta + r^2$ and $c_{\mathrm{QCS}}(\zeta,\alpha) = \kappa(\kappa-1) + \frac{3}{16}$.\par
We now turn to the classical analysis and consider SSS spacetimes admitting an horizon. They can be described by writing the metric in null Kruskal-Szekeres coordinates $(V,U,\phi,\theta)$ as
\begin{equation}\label{eq:sssmetric}
    ds^2_{\mathrm{SSS}} = \frac{f(\rho)}{\kappa^2 UV}\dd U \dd V + \rho^2 \dd \Omega_S,
\end{equation}
where the horizon $\rho= \rho_h$ is situated at the largest root of the function $f(\rho)$, $\kappa = \frac12 f'(r_h)$ is the surface gravity and $\dd \Omega_S$ is the area element of the unit 2-sphere. The radial coordinate is understood as a scalar field in the null coordinates $\rho= \rho(U,V)$. One can define a QCS dynamical system on the bifurcating surface $U=V=0$ by equipping a constant Schwarzschild-time slice $\Sigma_t$ with a symplectic structure---within the covariant phase space formalism---and taking the exterior of the horizon as the region of interest. The corner is then identified with the bifurcating horizon. The Hamiltonian vector fields are generated by the diffeomorphisms
\begin{Align}\label{eq:hamiltoniandiffeomorphisms}
    \xi_1 &= \frac12\qty(U \partial_U - V\partial V),\\
    \xi_2 &= \frac12(U \partial_V - V \partial_U),\\
    \xi_3 &= \frac12(U \partial_V + V \partial_U),\\
    \xi_4 &= \partial_U,\\
    \xi_5 &= \partial_V.
\end{Align}
For the specific metric \eqref{eq:sssmetric} and the corner being the bifurcating horizon, a straightforward calculation shows that only the boost generates a  non-vanishing Noether charge
\begin{equation}
    H_{\xi_1} = N^0_0 = \frac{\rho_h^2}{4}.
\end{equation}
In general there are many moment maps that associate the diffeomorphisms \eqref{eq:hamiltoniandiffeomorphisms} to $\mathfrak{ecs}$ generators such that the equivariance condition is respected. In a sense, this is because the moment map is inherently a local object which does not see the global structure of the underlying Lie group. Looking at $\xi_1,\xi_2,\xi_3$ as $\spl{2}$ generators however, it is clear that $\xi_2$ is the only compact transformation and should as such be associated with the eliptic generator of the special linear algebra. This in turn forces the following comoment map for the boost charge
\begin{equation}\label{eq:classicalmomentmap}
    \tilde{\mu}(D) = H_{\xi_1}.
\end{equation}
Note that the fact that Noether charge associated with the boost operator is related to the area of the corner was first discussed by Wald \cite{Wald:1993nt}.\par

According to the procedure described in the previous section, to connect this classical quantity to the quantum theory we must choose coherent states for which the Berezin symbols of the generators reproduce the classical (twisted) comoment map. This can be accomplished as follows. For each $\zeta\in\mathbb{D}$, fix $\alpha\in\mathbb{C}$ so that Eqs.~\eqref{eq:lH} and~\eqref{eq:lK} vanish\footnote{Here we choose an $\alpha \in \mathbb{C}$ that is independent of $c$ such that $l_{(\zeta,\alpha)\qty(X)} = l_{(\zeta,\alpha)\qty(P)} = 0$ in the classical limit $c\to 0$. One can however accommodate non-vanishing translation charges by choosing $\alpha$ to scale with $c^{-\frac12}$. This is in direct analogy with the standard treatment of Glauber coherent states in the quantum harmonic oscillator. There, the expectation value of the position and momentum operators scales like $\hbar^{\frac12}$ and one must held these classical values fixed when sending $\hbar$ to zero to obtain the classical limit.}. Substituting this value of $\alpha$ into Eq.~\eqref{eq:lD} and choosing $\theta=-\pi$ yields
\begin{equation}
    l_{(\zeta,\alpha)}(D) = \kappa - \frac14.
\end{equation}
The Berezin symbols of the $X,P$ charges vanish in the classical limit $c\to 0$, in agreement with the fact that the corresponding translation charges vanish. We may then identify the Berezin symbol of the operator $D$ with the classical moment map~\eqref{eq:classicalmomentmap}. Reinstating dimensions, the regime $\kappa\gg 1$ corresponds to large corners with respect to the Planck length $r_h >> \sqrt{\hbar G}$, and hence to the semiclassical limit. In this limit one can therefore identify the representation parameter $\kappa$ with the corner area in Planck units,
\begin{equation}
    \kappa \sim \frac{r_h^2}{G\hbar}, \qquad \kappa \gg 1.
\end{equation}
This identification was used in \cite{Ciambelli:2025ztm,Kowalski-Glikman:2025hom} to show that, for a suitable family of coherent states, the entanglement entropy between the spacetime regions inside and outside the horizon obeys an area law in the semiclassical limit.\par
Lastly, we comment on the coadjoint orbit describing the SSS spacetime. The twisted moment map for this field configuration gives
\begin{equation}
    \mu^{c}(g_{\mathrm{SSS}}) =  \frac{\rho_h^2}{4} d + c z
\end{equation}
where $d,z \in \mathfrak{qcs}^*$ are the dual vectors to the $D,Z$ generator. Using the modified Casimir function \eqref{eq:modifiedcasimir}, the corresponding coadjoint orbit of the ECS when $c\to0$ has vanishing Casimir $c_{\mathrm{ECS}} = 0$. It is interesting to note that moving the corner along the horizon away from the bifurcating surface generates a non-vanishing translation charge\footnote{More precisely, for a corner at $V=0$ with $U\neq 0$, the translation charge associated with $\partial_V$ is nonzero, whereas for a corner at $U=0$ with $V\neq 0$ the translation charge associated with $\partial_U$ is nonzero.}. Since on the horizon one always has either $U=0$ or $V=0$, at most one translation charge can be nonzero, and the corresponding ECS Casimir therefore remains zero. This suggests that the classical data associated with different corners are related by a coadjoint transformations, and hence---via the equivariance condition~\eqref{eq:equivariancecondition}---by field variations. One may therefore interpret the classical phase space of SSS gravity as the coadjoint orbit with vanishing ECS Casimir, with different points corresponding to different choices of corner on the horizon. We leave a detailed analysis of this perspective to future work.

\section{Conclusion}\label{sec:conclusion}
In this paper, we have shown how a quantum system described by the representations of the QCS group can be related to a classical system realizing the ECS symmetries through twisted moment maps on coadjoint orbits. We began with a review of the theory of coadjoint orbits and their twisted extensions, which serve as the geometric link between the quantum and classical descriptions. We then introduced Berezin symbols and showed that they are the Hamiltonian functions of the KKS form. This in turns allowed us to construct an explicit correspondence between the quantum and classical data by relating the Berezin symbols to twisted comoment map and then taking the twisting parameter to vanish to obtain the classical limit. We finally applied this formalism to gravity in static, spherically symmetric spacetimes which provides an example of QCS dynamical system. The correspondence allowed us to relate the representation parameter to the area of the corner under investigation.\par

A key strength of the corner proposal is that it provides a fully controlled Hilbert space for quantum gravity, constructed from symmetry considerations alone. At the same time, this feature underlies one of its main difficulties: in the absence of an \emph{a priori} notion of classical geometry, interpreting results derived purely from this structure can be challenging. The work presented in this paper lays the mathematical foundation for addressing this issue. In particular, classical geometry---and the associated observables---emerges only upon evaluating expectation values of quantum operators in coherent states.\par
From here, several directions deserve further investigation. First, the formalism developed here provides a sharper classical understanding of the corner proposal, thereby enabling one to address key questions in semiclassical gravity within the corner-proposal framework. For example, a straightforward extension to non-static settings would allow one to treat FRW spacetimes and could provide a fully quantum model of cosmology.\par
Another natural avenue is to generalize the construction to other QCS representations, in particular those induced by the principal series of the special linear group. This is, however, technically challenging: the results of the present work rely crucially on the K\"ahler structure of the unit disk, a feature that is absent for the one-sheeted hyperboloid coadjoint orbits of $\spl{2}$ associated with the principal series.\par
A further natural step is to develop the representation theory, coadjoint-orbit picture, and quantum--classical correspondence for higher-dimensional corner symmetry groups. While a full classification is likely to be extremely challenging, we expect that a tractable subset of representations can be constructed, and that this subset will already exhibit interesting physics.

% ---------------- Acknowledgments ----------------
\bigskip
\noindent\textbf{Acknowledgments.} I am deeply grateful to Jerzy Kowalski-Glikman for his supervision and encouragement throughout this project. I am also grateful to Giulio Neri for his help with several calculations, and for discussions that helped clarify a number of key points. I thank Luca Ciambelli for his guidance on the corner symmetry formalism and related aspects of this work. I am grateful to Rodrigo Andrade E Silva, Nicolas Cresto, Michael Imseis, Simon Langenscheidt, Leonardo Sanhueza Mardones and Simone Speziale for insightful discussions. I thank Perimeter Institute for its hospitality.
\par
\bigskip

\appendix
\section{Fundamental vector fields}\label{ap:vectorfields}
Fix $X\in\mathfrak{g}$ and let $\mathcal{X}$ be the fundamental vector field on $\Gamma$ generated by $X$.
To show $d\mu_\phi(\mathcal{X}_\phi)=X^\#_{\mu(\phi)}$, it suffices to test both vectors on the
linear coordinate functions $\chi_Y(m)=\langle m,Y\rangle$ on $\mathfrak{g}^*$, since the differentials
$d\chi_Y|_{\mu(\phi)}$ spans the whole cotangent space at $\mu(\phi)$. We have
\begin{equation*}
(\mu_* \mathcal{X})\chi_Y = 
\mathcal{X}\big(\chi_Y\circ\mu\big)
\;=\;
\mathcal{X}\big(\tilde\mu(Y)\big).
\end{equation*}
where we used the definition of the coordinate function to write $\chi_Y\circ \mu = \tilde{\mu}(Y)$. Using the definition of the comoment map and its equivariance, we can further write
\begin{equation*}
\mathcal{X}\big(\tilde\mu(Y)\big)
\;=\;
\{\tilde\mu(Y),\tilde\mu(X)\}_\Gamma
\;=\;
\tilde\mu([Y,X])
\;=\;
-\,\chi_{[X,Y]}\!\circ\mu .
\end{equation*}
On the other hand, for the coadjoint fundamental field one has
\begin{equation*}
X^\#(\chi_Y)(m)
\;=\;
\frac{d}{dt}\Big|_{t=0}\chi_Y\!\big(\mathrm{Ad}^*_{e^{tX}}m\big)
\;=\;
\langle \mathrm{ad}^*_X m, Y\rangle
\;=\;
\langle m,[Y,X]\rangle
\;=\;
-\,\chi_{[X,Y]}(m).
\end{equation*}
Evaluating at $m=\mu(\phi)$ shows
\begin{equation*}
X^\# \chi_Y = (\mu_*\mathcal{X})\chi_Y.
\end{equation*}
Since this holds for any $Y \in \mathfrak{g}$, we conclude $X^\# = \mu_* \mathcal{X}$.

\section{Classical Casimir transformation}\label{ap:casimir}
In the section we prove that any different choice of moment map, simply rescales the classical Casimir \eqref{eq:classicalcasimir} by an overall factor.
We start by considering the moment map $\mu^a_b$. Since any choice must satisfy the equivariance condition \eqref{eq:momentmapequivariance}, changing moment map corresponds to a automorphisms on the $\mathfrak{sl}\qty(2,\mathbb{R})$ algebra. They can be represented by $\spl{2}$ matrices $B^a_b$ with the moment map transforms as
\begin{equation}
    \mu'^a_b = B^a_c \mu^c_d \qty(B^{-1})^d_b.
\end{equation}
It is easy to see that the classical $\spl{2}$ Casimir is invariant under such a transformation
\begin{Align}
    C'^{\mathrm{cl}}_{\spl{2}} &= \mu'^a_b \mu'^b_a\\
    &= B^a_c (B^{-1})^d_b B^b_e (B^{-1})^n_a \mu^c_d \mu^e_n\\
    &= \mu^a_b \mu^b_a = C^{\mathrm{cl}}_{\spl{2}}.
\end{Align}
Let us now consider the Heisenberg part. It is well known that the group of automorphisms of the Heisenberg algebra, that is the group acting on the $x$ and $p$ that preserves the canonical commutation relation, is $\spl{2}$. However one can also consider $\mathrm{GL}\qty(2,\mathbb{R})$ elements, as long as the central moment map is rescalled accordingly. More precisely, different moment map choices are related by the transformations
\begin{align}
    \mu'^a &= A^a_b \mu^b\\
    \mu'^z &= \det(A) \mu^z,
\end{align}
with $A \in \mathrm{GL}\qty(2,\mathbb{R})$. An additional constraint arises from the mixed commutation relations
\begin{align}
    \qty\big{\mu'^a,\mu'^c_d}_\Gamma &= \qty\big((AB^{-1})^a_d B^c_b - \tfrac{1}{2} \delta^c_d A^a_b) \mu^b\\
    &= \delta^a_d \mu'^c - \tfrac{1}{2} \delta^c_d \mu'^a\\
    &= \qty\big(\delta^a_d A^c_b - \tfrac{1}{2} \delta^c_d A^a_b)\mu^b.
\end{align}
By equating the coefficients inside the parentheses and examining the different index configurations, one readily finds that
\begin{equation}
    A = t\,B.
\end{equation}
Further taking the determinant on both sides of the above gives
\begin{equation}
    t = \pm \sqrt{\det A}.
\end{equation}
One can then compute how the classical ECS Casimir changes under a change of moment map
\begin{Align}
    C'^{cl}_{\mathrm{ECS}} &= \frac12 \epsilon_{ac}\mu'^c \mu'^a_b \mu'^b\\
    &= \frac12 \epsilon_{ac} A^c_d A^b_e B^a_m(B^{-1})_b^l \mu_l^m \mu^e\mu^d\\
    &= \frac12 \qty(B^{-1}A)^l_e \epsilon_{mj} (B^{-1}A)^j_d \mu^m_l \mu^e \mu^d\\
   &= \frac12 \det(A) \epsilon_{ac}\mu^a_b\mu^b \mu^c\\
   &= \det(A)\, C^{cl}_{\mathrm{ECS}},
\end{Align}
where going from the second to the third line we used the fact that $B^T \epsilon = \epsilon B^{-1}$ for any matrix in $\spl{2}$. Putting this result together with the invariance of the $\spl{2}$ Casimir and the transformation of the moment map of the central element, we find
\begin{equation}
    C'^{\mathrm{cl}}_{\mathrm{QCS}} = C^{\mathrm{cl}}_{\mathrm{QCS}}.
\end{equation}
That is, the classical Casimir is invariant under transformation of the moment maps.
\section{Berezin functions are coordinate functions}\label{ap:berezinequalcoordinate}
To clarify the relationship between the Berezin functions and the coordinate function, let us note that the QCS coadjoint orbits admit a Kähler embedding in the projective Hilbert space through the coherent states\footnote{This follows from the fact that both the $\spl{2}$ and Heisenberg orbits admit Kähler embeddings~\cite{dey2022coadjointorbitskahlerstructure}, and that the QCS orbits were shown in~\cite{Neri:2025fsh} to be tensor products of these two.
} \cite{Onofri_75,coherentstateandkahlermanifolds}
\begin{equation}
    \Phi : G/H_m \longrightarrow \mathbb{P}\qty(\mathcal{H}^{s,c}), \quad [g]\mapsto \Phi(\qty[g]) = \qty[U([g]) \ket{\Omega}]\defeq \qty[\ket{\psi_g}].
\end{equation}
Note that the right hand side is nothing but the equivalence class of coherent states as defined in \eqref{eq:coherentstatedef}.
where the brackets denote the ray, i.e.\ the equivalence class of states differing by an overall phase and where $m\in \mathcal{O}^{(s,c)}$. 
This map is known as the Kodaira embedding.
The former being Kähler means that the Fubini-Study 2-form pulls back to the KKS form on the orbits \cite{FABerezin_1975,coherentstateandkahlermanifolds,RAWNSLEY199045,Bordemann:1993zv,Schlichenmaier_2010},
that is\footnote{Note that the Fubini-Study 2-form can be written as the exterior derivative of the Berry connection
\begin{equation}
  \Omega_{\mathrm{FS}} = \dd A \defeq -\dd\qty(\mel{\Omega}{\Theta}{\Omega})   
\end{equation}
Equation \eqref{eq:KKSequalFS} then follows from the Maurer-Cartan structure equation
\begin{equation}
    \dd \Theta + \Theta \wedge \Theta = 0.
\end{equation}
}
\begin{equation}\label{eq:KKSequalFS}
    \tau^*\Omega_{KKS} =  \mel{\Omega}{\Theta_\sigma \wedge \Theta_\sigma}{\Omega},
\end{equation}
where $\Theta$ is the left Maurer-Cartan form pulled back to the cosets by the displacement operator
\begin{equation}
    \Theta_\sigma = U_\sigma^{-1} dU_\sigma
\end{equation}
Note that although the pullback of the Maurer–Cartan form depends on the choice of representative $\sigma$, the KKS form does not, since the stabilizer subgroup acts only by a phase on the reference state.
Let us now contract the above with the fundamental vector field on the cosets. The former is defined as the pushforward of the fundamental vector field on the orbits by the inverse of the orbit map \eqref{eq:orbitmap}
\begin{equation}
    \mathsf{X} \defeq \qty(\tau^{-1})_* X^\#.
\end{equation}
Using the left group action on the cosets
\begin{equation}
    \triangleright: G \times G/H \longrightarrow G/H, \quad (g_1,\qty[g_2]) \mapsto g_1 \triangleright \qty[g_2]  = \qty[g_1 g_2]
\end{equation}
one can write the action of $\mathsf{X}$ on a function $\mathsf{f}\in C^{\infty}(G/H)$
\begin{equation}
    (\mathsf{X}\mathsf{f})(\qty[g]) \defeq \dv{t}\mathsf{f}(\exp(t X)\triangleright \qty[g])\eval_{t=0}.
\end{equation}
In order to compute the contraction, for each element of the coset $\qty[g]\in G/H$, we choose a section $\sigma: G/H \rightarrow G$ and use the definition of the displacement operator \eqref{eq:displacementoperator}
\begin{Align}
    \Theta_\sigma(\mathsf{X})([g]) &= U_\sigma^{-1}\qty([g]) \dv{t}U_\sigma(\qty[e^{tX} g])\eval_{t=0}\\
    &= U_\sigma^{-1}\qty([g_0]) \dd \pi(X)  U_\sigma\qty([g_0]).\\
\end{Align}
We can now calculate the contraction of the KKS form with the fundamental vector field
\begin{Align}
    \iota_{\mathsf{X}}\Omega_{KKS} &= \mel{\Omega}{\iota_{\mathsf{X}}\qty(\Theta_\sigma \wedge \Theta_\sigma)}{\Omega}\\
    &= \mel{\Omega}{\qty[\iota_{\mathsf{X}}\Theta_\sigma,\Theta_\sigma]}{\Omega}\\
    &= \mel{\Omega}{U_\sigma^{-1}\dd \pi\qty(X) \dd U_\sigma + \dd U_\sigma^{-1} \dd \pi\qty(X) U_\sigma}{\Omega}\\
    &= \dd \mel{\Omega}{U_\sigma^{-1}\dd \pi\qty(X)U_\sigma}{\Omega}\\
\end{Align}
Choosing the particular section \eqref{eq:perelomovsection} and evaluating the above at the point $\qty[g]\in G/H$ gives
\begin{equation}
    \iota_{\mathsf{X}}\qty(\tau^*\Omega^{(\alpha,\zeta)}_{\mathrm{KKS}}) = \dd \mel{\alpha,\zeta}{\dd \pi\qty(X)}{\alpha,\zeta}.
\end{equation}
That is, the expectation value of the algebra operator in a coherent state is a hamiltonian map for the associated fundamental vector field
\begin{equation}
    \iota_\mathsf{X} \tau^*\qty(\Omega_{\mathrm{KKS}})= \iota_{X^\#}\Omega_{\mathrm{KKS}} = \dd l^X.
\end{equation}

\bibliographystyle{ieeetr} 
\bibliography{refs}

\end{document}